\documentclass[,article,submit,pdftex,moreauthors]{Definitions/mdpi}
\renewcommand{\linenumbers}{} 

\firstpage{1} 
\pubvolume{1}
\issuenum{1}
\articlenumber{0}
\pubyear{2025}
\copyrightyear{2025}
\datereceived{ } 
\daterevised{ } 
\dateaccepted{ } 
\datepublished{ } 
\hreflink{https://doi.org/} 

\usepackage{subcaption}

\newenvironment{subproof}[1][\proofname]{%
  \renewcommand{\qedsymbol}{$\blacksquare$}%
  \begin{proof}[#1]%
}{%
  \end{proof}%
}

\Title{Nonparametric FBST for Validating Linear Models}

\TitleCitation{Nonparametric FBST for Validating Linear Models}

\Author{Rodrigo F. L. Lassance $^{1,2}$*\orcidA{}, Julio M. Stern $^{3}$\orcidC{} and Rafael B. Stern $^{3,}$\orcidB{}}

\AuthorNames{Rodrigo F. L. Lassance, Julio M. Stern and Rafael B. Stern}

\AuthorCitation{Lassance, R.F.L.; Stern, J.M.; Stern, R.B.}

\address{%
$^{1}$ \quad Department of Statistics, Federal University of São Carlos, São Paulo, Brazil\\
$^{2}$ \quad Institute of Mathematics and Computer Sciences, University of São Paulo, São Paulo, Brazil\\
$^{3}$ \quad Institute of Mathematics and Statistics, University of São Paulo, São Paulo, Brazil}

\corres{Correspondence: rflassance@gmail.com}

\abstract{The Full Bayesian Significance Test (FBST) possesses many desirable aspects, such as dismissing the need for hypotheses to have positive prior probability and providing a measure of evidence against $H_0$. Still, few attempts have been made to bring the FBST to nonparametric settings, with the main drawback being the need to obtain the highest posterior density (HPD) in a function space. In this work, we use a Gaussian processes prior to derive the FBST for hypotheses of the type
\begin{equation*}
    H_0: g(\boldsymbol{x}) = \boldsymbol{b}(\boldsymbol{x})\boldsymbol{\beta}, \quad \forall \boldsymbol{x} \in \mathcal{X}, \quad \boldsymbol{\beta} \in \mathbb{R}^k,
\end{equation*}
where $g(\cdot)$ is the regression function, $\boldsymbol{b}(\cdot)$ is a vector of linearly independent linear functions---such as $\boldsymbol{b}(\boldsymbol{x}) = \boldsymbol{x}'$---and $\mathcal{X}$ is the covariates' domain. We also make use of pragmatic hypotheses to verify if the data might be compatible with a linear model when factors such as measurement errors or utility judgments are accounted for. This contribution extends the theory of the FBST, allowing its application in nonparametric settings and providing a procedure that easily tests if linear models are adequate for the data and that can automatically perform variable selection.
}


\newcommand{\FBST}{

\tikzset{every picture/.style={line width=0.75pt}} 

\begin{tikzpicture}[x=0.75pt,y=0.75pt,yscale=-1,xscale=1]

\draw   (31,78.67) .. controls (31,63.39) and (43.39,51) .. (58.67,51) -- (141.67,51) .. controls (156.95,51) and (169.33,63.39) .. (169.33,78.67) -- (169.33,161.67) .. controls (169.33,176.95) and (156.95,189.33) .. (141.67,189.33) -- (58.67,189.33) .. controls (43.39,189.33) and (31,176.95) .. (31,161.67) -- cycle ;
\draw   (191,78.33) .. controls (191,63.05) and (203.39,50.67) .. (218.67,50.67) -- (301.67,50.67) .. controls (316.95,50.67) and (329.33,63.05) .. (329.33,78.33) -- (329.33,162) .. controls (329.33,177.28) and (316.95,189.67) .. (301.67,189.67) -- (218.67,189.67) .. controls (203.39,189.67) and (191,177.28) .. (191,162) -- cycle ;
\draw   (351,78.33) .. controls (351,63.05) and (363.39,50.67) .. (378.67,50.67) -- (461.67,50.67) .. controls (476.95,50.67) and (489.33,63.05) .. (489.33,78.33) -- (489.33,161.67) .. controls (489.33,176.95) and (476.95,189.33) .. (461.67,189.33) -- (378.67,189.33) .. controls (363.39,189.33) and (351,176.95) .. (351,161.67) -- cycle ;
\draw  [fill={rgb, 255:red, 167; green, 167; blue, 167 }  ,fill opacity=0.34 ] (38.95,136.28) .. controls (29.9,123) and (47.51,95.24) .. (78.28,74.28) .. controls (109.06,53.32) and (141.33,47.1) .. (150.38,60.38) .. controls (159.43,73.67) and (141.82,101.43) .. (111.05,122.38) .. controls (80.28,143.34) and (48,149.57) .. (38.95,136.28) -- cycle ;
\draw  [fill={rgb, 255:red, 167; green, 167; blue, 167 }  ,fill opacity=0.34 ] (199.95,136.28) .. controls (190.9,123) and (208.51,95.24) .. (239.28,74.28) .. controls (270.06,53.32) and (302.33,47.1) .. (311.38,60.38) .. controls (320.43,73.67) and (302.82,101.43) .. (272.05,122.38) .. controls (241.28,143.34) and (209,149.57) .. (199.95,136.28) -- cycle ;
\draw  [fill={rgb, 255:red, 167; green, 167; blue, 167 }  ,fill opacity=0.34 ] (359.95,136.28) .. controls (350.9,123) and (368.51,95.24) .. (399.28,74.28) .. controls (430.06,53.32) and (462.33,47.1) .. (471.38,60.38) .. controls (480.43,73.67) and (462.82,101.43) .. (432.05,122.38) .. controls (401.28,143.34) and (369,149.57) .. (359.95,136.28) -- cycle ;
\draw  [fill={rgb, 255:red, 208; green, 2; blue, 27 }  ,fill opacity=0.46 ] (411,155.5) .. controls (411,141.69) and (422.19,130.5) .. (436,130.5) .. controls (449.81,130.5) and (461,141.69) .. (461,155.5) .. controls (461,169.31) and (449.81,180.5) .. (436,180.5) .. controls (422.19,180.5) and (411,169.31) .. (411,155.5) -- cycle ;
\draw  [draw opacity=0][fill={rgb, 255:red, 255; green, 0; blue, 0 }  ,fill opacity=1 ][line width=0.75]  (433.05,155.5) .. controls (433.05,153.87) and (434.37,152.55) .. (436,152.55) .. controls (437.63,152.55) and (438.95,153.87) .. (438.95,155.5) .. controls (438.95,157.13) and (437.63,158.45) .. (436,158.45) .. controls (434.37,158.45) and (433.05,157.13) .. (433.05,155.5) -- cycle ;
\draw  [fill={rgb, 255:red, 208; green, 2; blue, 27 }  ,fill opacity=0.46 ] (91,105.5) .. controls (91,91.69) and (102.19,80.5) .. (116,80.5) .. controls (129.81,80.5) and (141,91.69) .. (141,105.5) .. controls (141,119.31) and (129.81,130.5) .. (116,130.5) .. controls (102.19,130.5) and (91,119.31) .. (91,105.5) -- cycle ;
\draw  [draw opacity=0][fill={rgb, 255:red, 255; green, 0; blue, 0 }  ,fill opacity=1 ][line width=0.75]  (113.05,105.5) .. controls (113.05,103.87) and (114.37,102.55) .. (116,102.55) .. controls (117.63,102.55) and (118.95,103.87) .. (118.95,105.5) .. controls (118.95,107.13) and (117.63,108.45) .. (116,108.45) .. controls (114.37,108.45) and (113.05,107.13) .. (113.05,105.5) -- cycle ;
\draw  [fill={rgb, 255:red, 208; green, 2; blue, 27 }  ,fill opacity=0.46 ] (251,129.5) .. controls (251,115.69) and (262.19,104.5) .. (276,104.5) .. controls (289.81,104.5) and (301,115.69) .. (301,129.5) .. controls (301,143.31) and (289.81,154.5) .. (276,154.5) .. controls (262.19,154.5) and (251,143.31) .. (251,129.5) -- cycle ;
\draw  [draw opacity=0][fill={rgb, 255:red, 255; green, 0; blue, 0 }  ,fill opacity=1 ][line width=0.75]  (273.05,129.5) .. controls (273.05,127.87) and (274.37,126.55) .. (276,126.55) .. controls (277.63,126.55) and (278.95,127.87) .. (278.95,129.5) .. controls (278.95,131.13) and (277.63,132.45) .. (276,132.45) .. controls (274.37,132.45) and (273.05,131.13) .. (273.05,129.5) -- cycle ;

\draw (38,166) node [anchor=north west][inner sep=0.75pt]   [align=left] {$\mathbb{H}$};
\draw (198,166) node [anchor=north west][inner sep=0.75pt]   [align=left] {$\mathbb{H}$};
\draw (358,166) node [anchor=north west][inner sep=0.75pt]   [align=left] {$\mathbb{H}$};
\draw (31.2,10) node [anchor=north west][inner sep=0.75pt]   [align=left] {\small $\bullet$ Do not reject $H_0$\\ \small $\bullet$ Do not reject $Pg(H_0)$};
\draw (191.2,10) node [anchor=north west][inner sep=0.75pt]   [align=left] {\small $\bullet$ Reject $H_0$\\ \small $\bullet$ Do not reject $Pg(H_0)$};
\draw (352.2,10) node [anchor=north west][inner sep=0.75pt]   [align=left] {\small $\bullet$ Reject $H_0$\\ \small $\bullet$ Reject $Pg(H_0)$};
\draw (55.75,71.51) node [anchor=north west][inner sep=0.75pt]  [rotate=-326.14] [align=left] {\textit{HPD}};
\draw (216.75,71.51) node [anchor=north west][inner sep=0.75pt]  [rotate=-326.14] [align=left] {\textit{HPD}};
\draw (376.75,71.51) node [anchor=north west][inner sep=0.75pt]  [rotate=-326.14] [align=left] {\textit{HPD}};
\draw (451.06,110.84) node [anchor=north west][inner sep=0.75pt]  [font=\small,rotate=-48.18] [align=left] {$Pg(H_0)$};
\draw (425,136.5) node [anchor=north west][inner sep=0.75pt]   [align=left] {$H_0$};
\draw (131.06,59.84) node [anchor=north west][inner sep=0.75pt]  [font=\small,rotate=-48.18] [align=left] {$Pg(H_0)$};
\draw (105,86.5) node [anchor=north west][inner sep=0.75pt]   [align=left] {$H_0$};
\draw (291.06,83.84) node [anchor=north west][inner sep=0.75pt]  [font=\small,rotate=-48.18] [align=left] {$Pg(H_0)$};
\draw (265,110.5) node [anchor=north west][inner sep=0.75pt]   [align=left] {$H_0$};

\end{tikzpicture}
}

\begin{document}


\section{Introduction}

Although linear models are widespread in the scientific literature, their validity is rarely tested in its full complexity. Generally, linearity is tested as a particular case of a more general parametric model \citep{kershaw2001} or compared to a finite selection of models---each with their own prior specification---through measures such as the Deviance Information Criterion \citep{spiegelhalter2002}. In actuality, testing the adherence of linear models to data requires \textbf{(i)} assigning a nonparametric prior to the set of regression functions and \textbf{(ii)} devising a procedure that highlights the evidence against the linear model hypothesis based on the data and the prior. Devising a test solely based on the posterior probability of the hypothesis in this case is seldom advised, as it imposes positive prior probability to the set of linear models when there are countless nonlinear functions arbitrarily close to any element of it.

The Full Bayesian Significance Test (FBST, \cite{pereira2008}) is the testing framework used throughout this work. The FBST does not violate the likelihood principle, does not require setting positive prior probabilities to hypotheses and provides a measure of evidence against $H_0$, along with other desirable characteristics. With the exception of \citet{liu2024}, the FBST has not been applied to nonparametrics \citep{pereira2020}, still requiring theoretical developments to systematically embrace such settings.

Bridging the gaps above, this paper provides a nonparametric FBST formulation that tests the adherence of linear models to data. By using a Gaussian Process (GP, \citep{rasmussen2006}) as a prior to model the regression function, we propose FBST procedures that depend on whether the covariates' domain $\mathcal{X}$ is finite or infinite. Furthermore, we lay out FBST procedures for hypotheses that include negligible deviations from $H_0$, known as pragmatic hypotheses \citep{esteves2019}, useful to evaluate if $H_0$ is approximately instead of precisely compatible with the data.

\begin{figure}[htb]
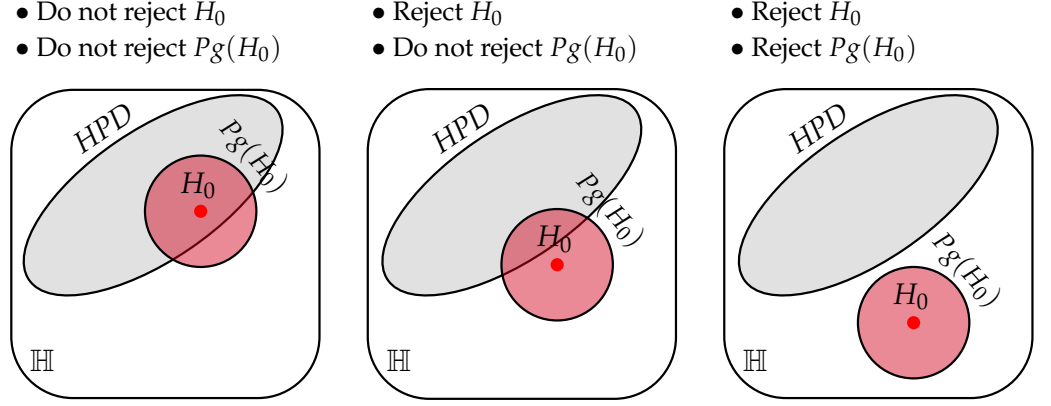

\begin{center}
    \resizebox{1\linewidth}{!}{\FBST}
    \caption{Illustration of the FBST for a precise $H_0$ and its pragmatic version $Pg(H_0)$ in the hypothesis space $\mathbb{H}$. Each panel presents a possible configuration of the hypotheses and the HPD, with the text above the panels indicating the conclusion.}
    \label{fig:fbst_illustration}
\end{center}
\end{figure}

In \autoref{fig:fbst_illustration}, we illustrate how the FBST operates when applied to $H_0$ and its pragmatic version, $Pg(H_0)$. For $\alpha \in (0, 1)$, the posterior is used to obtain the $(1 - \alpha)100\%$ Highest Posterior Density (HPD) region, the smallest credible region with probability $(1 - \alpha)$ of containing the quantity of interest. The hypothesis is rejected if it does not intersect with the HPD. This procedure is meaningful even when $H_0$ is precise, that is, when $\mathbb{P}(H_0) = 0$.

Even though our contribution makes exclusive use of the FBST, this does not imply that it is the only valid framework for the problem. While no other testing procedure as general as ours has been proposed, \citet{mulder2023} uses Bayes factors to test if a single covariate may be nonlinearly related to the response variable and \citet[][section 3.1]{lassance2025} tests the pragmatic version of the linear model hypothesis through its posterior probability.

This work is organized as follows. In \autoref{sct:methods}, the required background knowledge is provided. Our findings are presented in \autoref{sct:res}, leading in \autoref{sct:droplet} to an application that puts all the FBST procedures to use. Lastly, \autoref{sct:disc} describes how to enhance the FBST further and establishes potential future research. All proofs can be found in \autoref{sct:proof}.

\section{Materials and Methods}
\label{sct:methods}

\subsection{Full Bayesian Significance Test (FBST)}
The FBST is composed of three steps \citep{pereira2008}. For $H_0: \theta \in \Theta_0 \subset \Theta$, where $\Theta$ is the parameter space, these steps are:
\begin{enumerate}
    \item Delimit the set of elements in $\Theta$ that are more likely than those in $\Theta_0$. That is, if $f(\theta|\mathcal{D})$ is the posterior density of $\theta$ given the data $\mathcal{D}$,
    $$
        T := \left\{\theta \in \Theta: f(\theta|\mathcal{D}) \ge \sup_{\theta \in \Theta_0} f(\theta|\mathcal{D})\right\}.
    $$
    \item Obtain $\text{e-value} := 1 - \mathbb{P}(\theta \in T | \mathcal{D}) = 1 - \int_{T} f(\theta | \mathcal{D}) d\theta,$ the Bayesian evidence value.
    \item Reject $H_0$ if $\text{e-value} \le \alpha$ for a previously specified significance level $\alpha \in (0,1)$.
\end{enumerate}

In this paper, we use a procedure equivalent to the FBST: reject $H_0$ if the Highest Posterior Density (HPD) region is such that $HPD \cap \Theta_0 = \emptyset$. The HPD is the smallest region with posterior probability of $1 - \alpha$, obtained by finding the value $f^*$ such that
$$
    \mathbb{P}(\theta \in HPD | \mathcal{D}) = 1 - \alpha, \quad HPD := \left\{\theta \in \Theta: f(\theta|\mathcal{D}) \ge f^*\right\}.
$$
When $\theta|\mathcal{D}$ is normally distributed, the HPD region is equivalent to the credible interval symmetric around the posterior mean. For its multivariate counterpart, $\boldsymbol{\theta}|\mathcal{D} \sim N_k(\boldsymbol{\mu}, \Sigma)$, we have that $\Sigma^{-1/2} (\boldsymbol{\theta} - \boldsymbol{\mu}) \sim N_k(\boldsymbol{0}, \mathbb{I})$, and thus $(\boldsymbol{\theta} - \boldsymbol{\mu})' \Sigma^{-1} (\boldsymbol{\theta} - \boldsymbol{\mu}) \sim \chi^2_{k}$, where $\chi^2_{k}$ stands for the chi-squared distribution with $k$ degrees of freedom. Therefore, if $q_{(1 - \alpha)}(\cdot)$ is the $(1 - \alpha)$100\% quantile function,
the HPD is given by the following ellipsoid \citep[][Result 4.7]{johnson2002}:
\begin{equation}
    \label{eq:mnorm_ci}
    \{ \boldsymbol{\theta} \in \mathbb{R}^k: (\boldsymbol{\theta} - \boldsymbol{\mu})' \Sigma^{-1} (\boldsymbol{\theta} - \boldsymbol{\mu}) \le q_{(1 - \alpha)}(\chi^2_{k})\}.
\end{equation}

\subsection{Gaussian Processes (GP)}
\label{sct:gp}

A GP is a nonparametric family of priors used to model functions in regression settings. The random function $g: \mathcal{X} \rightarrow \mathbb{R}$ behaves according to a GP if
    $$
        g(\boldsymbol{X}) \sim N(m(\boldsymbol{X}), K(\boldsymbol{X}, \boldsymbol{X})), \quad \forall \boldsymbol{X} \subset \mathcal{X},
    $$
where $m(\cdot)$ and $K(\cdot, \cdot)$ respectively determine the mean and covariance of the process. When the response variable $Y$ is such that $Y = g(\boldsymbol{x}) + \epsilon$ for $\epsilon \sim N(0, \sigma^2)$, that is,
$$
    L(g, \sigma^2 | \boldsymbol{y}, \boldsymbol{X}) = (2\pi \sigma^2)^{-n/2}\exp\left\{-\frac{1}{2\sigma^2}\sum_{i=1}^n\Big(y_i - g(\boldsymbol{x}_i)\Big)^2\right\},
$$
then the GP is conjugate and its posterior is such that

\begin{align*}
    g(\boldsymbol{X'})&|\boldsymbol{y}, \boldsymbol{X}, \sigma^2 \sim N(\mu(\boldsymbol{X}'), \Sigma(\boldsymbol{X}', \boldsymbol{X}')),\\
    \mu(\boldsymbol{X}') &:= m(\boldsymbol{X}') + K(\boldsymbol{X}, \boldsymbol{X}')(K(\boldsymbol{X}, \boldsymbol{X}) + \sigma^2 \mathbb{I})^{-1}(\boldsymbol{y} - m(\boldsymbol{X})),\\
    \Sigma(\boldsymbol{X}', \boldsymbol{X}') &:= K(\boldsymbol{X}', \boldsymbol{X}') - K(\boldsymbol{X}, \boldsymbol{X}') (K(\boldsymbol{X}, \boldsymbol{X}) + \sigma^2\mathbb{I})^{-1} K(\boldsymbol{X}', \boldsymbol{X}).
\end{align*}

The choice of $m$, $K$ and $\sigma^2$ reflect positions on the mean, smoothness and variation surrounding the GP. In \autoref{sct:droplet}, we use the specifics of the application to choose them. For more general settings, one may assume that the uncertainty of $m$ and $K$ is reducible to a finite number of parameters. Then, one can either set priors to such parameters directly \citep{oakley2002} or plug point estimates for them based on the maximum partial likelihood \citep{wang2016}.

Conditionally on $\sigma^2$, the HPD region of the GP can be analytically obtained for any finite set $\boldsymbol{X}' = (\boldsymbol{x}_1, \boldsymbol{x}_2, \cdots, \boldsymbol{x}_m)'$. Since the marginals of the posterior GP are also normally distributed, \autoref{eq:mnorm_ci} entails that the $(1 - \alpha)100\%$ HPD region for $g(\boldsymbol{X}')|\boldsymbol{y}, \boldsymbol{X}$ is
\begin{equation}
    \label{eq:gp_ci}
    \{ h(\boldsymbol{X}') \in \mathbb{R}^m: (h(\boldsymbol{X}') - \mu(\boldsymbol{X}'))' \Sigma(\boldsymbol{X}', \boldsymbol{X}')^{-1} (h(\boldsymbol{X}') - \mu(\boldsymbol{X}')) \le q_{(1 - \alpha)}(\chi^2_{m})\}.
\end{equation}

It is also possible to obtain an HPD set for the GP without setting $\boldsymbol{X}'$. Let $\mathbb{P}_{g}$ and $\mathbb{P}_{g|\boldsymbol{y}, \boldsymbol{X}, \sigma^2}$ respectively be the prior and posterior probability measures of the GP defined on a measurable space $(\mathcal{G}, \mathbb{G})$. Hence,
$$
    \mathbb{P}_{g|\boldsymbol{y}, \boldsymbol{X}, \sigma^2}(A) = \dfrac{\int_{A} L(g, \sigma^2 | \boldsymbol{y}, \boldsymbol{X})d\mathbb{P}_{g}(h)}{\int_{\mathcal{G}} L(g, \sigma^2 | \boldsymbol{y}, \boldsymbol{X}) d\mathbb{P}_{g}(h)}, \quad \forall A \subset \mathbb{G}.
$$
Since $\mathbb{P}_{g|\boldsymbol{y}, \boldsymbol{X}, \sigma^2} \ll \mathbb{P}_{g}$, the Radon-Nikodym derivative of the GP for $h \in \mathcal{G}$ is such that
\begin{equation}
    \label{eq:rn_gp}
    \frac{d\mathbb{P}_{g|\boldsymbol{y},\boldsymbol{X}, \sigma^2}}{d\mathbb{P}_g}(h) \propto \exp\left(-\frac{1}{2\sigma^2}\sum_{i = 1}^n(y_i - h(\boldsymbol{x}_i))^2\right) = \exp\left(-\frac{1}{2\sigma^2}(\boldsymbol{y} - h(\boldsymbol{X}))'(\boldsymbol{y} - h(\boldsymbol{X}))\right),
\end{equation}
i.e., it suffices to evaluate $h$ only on the values of $\boldsymbol{X}$ in the sample. To account for repeated lines in $\boldsymbol{X}$, let $\boldsymbol{X}^*$ be the matrix with all unique observations from $\boldsymbol{X}$ and $n^*$ be the number of lines of $\boldsymbol{X}^*$. Defining $D_{n_{\boldsymbol{X}^*}}$ as a diagonal matrix that counts how many times each $\boldsymbol{x} \in \boldsymbol{X}^*$ appears in $\boldsymbol{X}$ and $\overline{\boldsymbol{y}}_{\boldsymbol{X}^*}$ as the vector of means of all elements of $\boldsymbol{y}$ related to each $\boldsymbol{x}$,
\begin{equation}
    \label{eq:rn_final}
    \frac{d\mathbb{P}_{g|\boldsymbol{y},\boldsymbol{X}, \sigma^2}}{d\mathbb{P}_g}(h) \propto \exp\left(-\frac{1}{2\sigma^2}(h(\boldsymbol{X}^*) - \overline{\boldsymbol{y}}_{\boldsymbol{X}^*})'D_{n_{\boldsymbol{X}^*}}(h(\boldsymbol{X}^*) - \overline{\boldsymbol{y}}_{\boldsymbol{X}^*})\right).
\end{equation}

Thus, for a constant $c_\alpha$, a Weighted Residual Sum of Squares (WRSS) defines the HPD:
{\small
\begin{equation}
    \label{eq:hpd_gp}
    HPD_{(1-\alpha)} = \left\{h \in \mathcal{G}: WRSS(h) \le c_\alpha\right\}, WRSS(h) := (h(\boldsymbol{X}^*) - \overline{\boldsymbol{y}}_{\boldsymbol{X}^*})'D_{n_{\boldsymbol{X}^*}}(h(\boldsymbol{X}^*) - \overline{\boldsymbol{y}}_{\boldsymbol{X}^*}).
\end{equation}
}
\subsection{Pragmatic hypotheses}

The pragmatic hypothesis enlarges $H_0$ to a set deemed as practically equivalent. The implementation
uses the notion of negligible deviations from $H_0$. The degree to which the hypothesis is enlarged depends on the choice of a threshold $\varepsilon$, and factors such as the scale of measurement errors or expert's utility judgments could help set it (see \autoref{sct:droplet} for a practical example and \citet[][section 4]{lassance2025} for suggestions). Formally, for a hypothesis space $\mathbb{H}$, let $d(\cdot,\cdot)$ be the dissimilarity function from which one can express how much of a departure from $H_0$ is reasonable. Then, the pragmatic hypothesis is given by
\begin{equation}
    Pg(H_0, d, \varepsilon) := \bigcup_{h_0 \in H_0}\left\{h \in \mathbb{H}: d(h_0,h) \le \varepsilon\right\} = \left\{h \in \mathbb{H}: \inf_{h_0 \in H_0} d(h_0,h) \le \varepsilon \right\},
    \label{eq:prag_hyp}
\end{equation}
that is, the pragmatic hypothesis contains all elements $h \in \mathbb{H}$ such that, for some element $h_0 \in H_0$, $d(h_0, h) \le \varepsilon$. In this work, we assume that $\mathbb{H} = \mathcal{G}$ is a space of functions of the type $h: \mathcal{X} \rightarrow \mathbb{R}$. Further specifications on $\mathbb{H}$ are presented in \autoref{sct:res}. When $d(\cdot, \cdot)$ and $\varepsilon$ are implicit, we use $Pg(H_0)$ to denote the pragmatic hypothesis.

\section{Results}
\label{sct:res}
Throughout this work, we use the modeling assumptions in \autoref{sct:gp} for the data $(\boldsymbol{y, \boldsymbol{X}})$ and the regression function $g(\cdot)$, and assume that the hypothesis of interest is
\begin{equation}
    \label{eq:lin_hyp}
    H_0: g(\boldsymbol{x}) = \boldsymbol{b}(\boldsymbol{x})\boldsymbol{\beta}, \quad \forall \boldsymbol{x} \in \mathcal{X}, \quad \boldsymbol{\beta} \in \mathbb{R}^k,
\end{equation}
where $\boldsymbol{b}(\boldsymbol{x}) = (b_1(\boldsymbol{x}), b_2(\boldsymbol{x}), \cdots, b_k(\boldsymbol{x})) \subset \mathbb{H}$ is a linearly independent set of linear functions and $\mathcal{X}$ is the covariates' domain. The choice of $\boldsymbol{b}$ determines the test performed, such as evaluating linear models ($\boldsymbol{b}(\boldsymbol{x}) := \boldsymbol{x}$) or doing variable selection ($\boldsymbol{b}(\boldsymbol{x}) := \boldsymbol{x}_{-i}$).

Our findings are divided in two settings: those applicable to $H_0$ and those to $Pg(H_0)$. In both cases, we explore when $\mathcal{X}$ is a finite or an infinite set. The finite case provides a closed-form solution for the FBST of $H_0$ and a solution for the pragmatic hypothesis that requires a univariate optimization procedure. When $\mathcal{X}$ is infinite, testing $H_0$ or $Pg(H_0)$ also requires determining $c_\alpha$ in the HPD of \autoref{eq:hpd_gp}, which is achieved by noting that the $WRSS$ can be expressed as a linear combination of noncentral chi-squared random variables, therefore $c_\alpha$ is the $(1-\alpha)$ quantile of a generalized chi-squared distribution \citep{davies1980}.

\begin{Theorem}[FBST of the linear model hypothesis]
    \label{thm:lm_hpd}
    Let $H_0$ be the hypothesis in \autoref{eq:lin_hyp} and $g(\cdot)|\boldsymbol{y}, \boldsymbol{X} \sim GP(\mu(\cdot), \Sigma(\cdot, \cdot))$. Then,

    \begin{itemize}
        \item When $\mathcal{X}$ is a finite set, the FBST does not reject $H_0$ if and only if
        \begin{equation*}
            \left(\boldsymbol{b}(\mathcal{X})\hat{\boldsymbol{\beta}} - \mu(\mathcal{X})\right)' \Sigma(\mathcal{X}, \mathcal{X})^{-1} \left(\boldsymbol{b}(\mathcal{X})\hat{\boldsymbol{\beta}} - \mu(\mathcal{X})\right) \le q_{(1 - \alpha)}(\chi^2_{|\mathcal{X}|}),
        \end{equation*}
        where $\hat{\boldsymbol{\beta}} = \left(\boldsymbol{b}(\mathcal{X})'\Sigma(\mathcal{X}, \mathcal{X})^{-1}\boldsymbol{b}(\mathcal{X})\right)^{-1} \boldsymbol{b}(\mathcal{X})'\Sigma(\mathcal{X}, \mathcal{X})^{-1}\mu(\mathcal{X})$ and $|\mathcal{X}|$ is the size of $\mathcal{X}$.

        \item When $\mathcal{X}$ is an infinite set, the FBST does not reject $H_0$ if and only if
        \begin{align*}
            &\boldsymbol{y}'\boldsymbol{M}\boldsymbol{y} \le c_\alpha, \quad \boldsymbol{M} = \mathbb{I} - \boldsymbol{b}(\boldsymbol{X})(\boldsymbol{b}(\boldsymbol{X})'\boldsymbol{b}(\boldsymbol{X}))^{-1}\boldsymbol{b}(\boldsymbol{X})'.
        \end{align*}
    \end{itemize}
\end{Theorem}

Before presenting the FBST for the pragmatic version of \autoref{eq:lin_hyp}, we specify $\mathbb{H}$ and provide the infimum when the dissimilarity function in \autoref{eq:prag_hyp} is the $L^2$ distance in the probability space of $\boldsymbol{X}$. The hypothesis space $\mathbb{H}$ is such that
\begin{equation}
    \label{eq:hyp_space}
    h \in \mathbb{H} \Longleftrightarrow \mathbb{E}_{\boldsymbol{X}}(h^2) = \int_{\mathcal{X}} h(\boldsymbol{x})^2 d\mathbb{P}(\boldsymbol{x}) < \infty.
\end{equation}
As for the infimum, it is described in the following Lemma:
\begin{Lemma}[Infimum of the dissimilarity on the linear model set]\label{lem:lm_inf}
    Let \autoref{eq:hyp_space} denote the hypothesis space and $H_0$ be the hypothesis in \autoref{eq:lin_hyp}. If $d(h_0, h) := \sqrt{\mathbb{E}_{\boldsymbol{X}}[(h_0 - h)^2]}$, then  $d(H_0, h) = d(\boldsymbol{b} \times \tilde{\boldsymbol{\beta}}_h, h), \forall h \in \mathbb{H}$, where
    \begin{align*}
        \tilde{\boldsymbol{\beta}}_h = A_{\boldsymbol{b}}^{-1} \times \boldsymbol{h}_{\boldsymbol{b}},
        \quad A_{\boldsymbol{b}} &=
        \left(\begin{array}{cccc}
             \mathbb{E}[b_1^2(\boldsymbol{X})] & \mathbb{E}[b_2(\boldsymbol{X}) b_1(\boldsymbol{X})] & \cdots & \mathbb{E}[b_k(\boldsymbol{X}) b_1(\boldsymbol{X})] \\
             \mathbb{E}[b_1(\boldsymbol{X}) b_2(\boldsymbol{X})] & \mathbb{E}[b_2^2(\boldsymbol{X})] & \cdots & \mathbb{E}[b_k(\boldsymbol{X}) b_2(\boldsymbol{X})] \\
             \vdots     & \vdots     & \ddots & \vdots     \\
             \mathbb{E}[b_1(\boldsymbol{X}) b_k(\boldsymbol{X})] & \mathbb{E}[b_2(\boldsymbol{X}) b_k(\boldsymbol{X})] & \cdots & \mathbb{E}[b_k^2(\boldsymbol{X})]
        \end{array}\right),\\
        \boldsymbol{h}_{\boldsymbol{b}}' &= \Big(\mathbb{E}[h(\boldsymbol{X}) b_1(\boldsymbol{X})], \quad \mathbb{E}[h(\boldsymbol{X}) b_2(\boldsymbol{X})], \quad \cdots \quad, \quad \mathbb{E}[h(\boldsymbol{X}) b_k(\boldsymbol{X})] \Big).
    \end{align*}
\end{Lemma}

\begin{Theorem}[FBST of the pragmatic version of $H_0$]
    \label{thm:pg_hpd}
    Let $\mathbb{H}$ be given by \autoref{eq:hyp_space} and define $d(h_0, h) := \sqrt{\mathbb{E}_{\boldsymbol{X}}[(h_0 - h)^2]}$. Assume that $\sum_{\boldsymbol{x} \in \boldsymbol{X}^*}\mathbb{P}(\boldsymbol{x}) > 0$. Then,
    \begin{itemize}
        \item When $\mathcal{X}$ is a finite set, the FBST does not reject $Pg(H_0)$ if and only if
        $$
        \exists s \in (0,1) : \quad 1 - \mu(\mathcal{X})'\left(\frac{\varepsilon^2}{1-s}\boldsymbol{N}^{-1} + \frac{1}{s}\Sigma(\mathcal{X}, \mathcal{X})q_{(1 - \alpha)}(\chi^2_{|\mathcal{X}|}) \right)\mu(\mathcal{X})< 0,
        $$
        where $\boldsymbol{N} := D_{\mathbb{P}(\mathcal{X})}\left[\mathbb{I} - \boldsymbol{b}(\mathcal{X})\left(\boldsymbol{b}(\mathcal{X})'D_{\mathbb{P}(\mathcal{X})}\boldsymbol{b}(\mathcal{X})\right)^{-1}\boldsymbol{b}(\mathcal{X})'D_{\mathbb{P}(\mathcal{X})}\right]$ and $D_{\mathbb{P}(\mathcal{X})}$ is a diagonal matrix formed by the vector $\mathbb{P}(\boldsymbol{x}), \boldsymbol{x} \in \mathcal{X}$.

        \item When $\mathcal{X}$ is an infinite set, the FBST does not reject $Pg(H_0)$ if and only if
        $$
            \exists s \in (0,1) : \quad 1 - \overline{\boldsymbol{y}}_{\boldsymbol{X}^*}'\left(\frac{\varepsilon^2}{1-s}\boldsymbol{M}^{-1} + \frac{1}{s}D_{n_{\boldsymbol{X}^*}}c_\alpha \right)\overline{\boldsymbol{y}}_{\boldsymbol{X}^*}< 0,
        $$
        where $\boldsymbol{M} := D_{\mathbb{P}(\boldsymbol{X}^*)}\left[\mathbb{I} - \boldsymbol{b}(\boldsymbol{X}^*)\left(\boldsymbol{b}(\boldsymbol{X}^*)'D_{\mathbb{P}(\boldsymbol{X}^*)}\boldsymbol{b}(\boldsymbol{X}^*)\right)^{-1}\boldsymbol{b}(\boldsymbol{X}^*)'D_{\mathbb{P}(\boldsymbol{X}^*)}\right]$ and $D_{\mathbb{P}(\boldsymbol{X}^*)}$ is a diagonal matrix formed by the vector $\mathbb{P}(\boldsymbol{x}), \boldsymbol{x} \in \boldsymbol{X}^*$.
    \end{itemize}
\end{Theorem}

In the infinite case of $\mathcal{X}$ in \autoref{thm:pg_hpd}, an appealing choice for the distribution of $\boldsymbol{X}$ is based on the posterior of a Dirichlet process \citep{ferguson1973}. For a concentration parameter $\tau$ and a centering distribution $\pi$, set a Dirichlet process $P \sim DP(\tau, \pi)$ such that $\boldsymbol{X} | P \sim P$. Then,
$$
    P | \boldsymbol{x} \sim DP\left(\frac{\tau\pi}{\tau + n} + \frac{\sum_{i = 1}^n \delta_{\boldsymbol{x}_i}}{\tau + n}, \tau + n\right) \Longrightarrow \mathbb{P}(\boldsymbol{X}_{new} \in A | \boldsymbol{X} = \boldsymbol{x}) = \frac{\tau\pi(A)}{\tau + n} + \frac{\sum_{i = 1}^n \mathbb{I}(\boldsymbol{x}_i \in A)}{\tau + n},
$$
where $\delta_{\boldsymbol{x}_i} = \mathbb{I}(\boldsymbol{x}_i \in A)$ for $A \subseteq \mathcal{X}$. With this choice, one can ensure that positive probability will always be assigned to all $\boldsymbol{x} \in \boldsymbol{X}^*$. Moreover, $\tau$ can leverage the weight of the prior on the FBST, with higher values of $\tau$ leading to a higher chance of not rejecting $Pg(H_0)$.

\section{Application: Water droplet experiment}
\label{sct:droplet}

The dataset from \citet{duguid1969} provides a setting where small water droplets (ranging from 3 to 9 micrometers) are free falling through a tube that keeps factors such as temperature and humidity constant. As a droplet falls, a camera takes a picture at every 0.5 second, ceasing activity after $7$ seconds. One of the objectives of the study was to evaluate Fick's law, which in this setting implies that---when time is a covariate---the decrease in radius of the droplet can be described through a linear model. The two hypotheses of interest are
$$
    \left\{\begin{array}{l}
        H_0^1: g(t) = \beta_0 + \beta_1 t, \quad \forall t \in \{0s, 0.5s, \cdots, 7s\}, \quad (\beta_0, \beta_1) \in \mathbb{R}^2;\\
        H_0^2: g(t) = \beta_0, \quad \forall t \in \{0s, 0.5s, \cdots, 7s\}, \quad \beta_0 \in \mathbb{R};
        \end{array}\right.
$$
with the first hypothesis testing the validity of Fick's law for this case and the second one verifying if time can be removed as a covariate.

\begin{figure}[ht]
    \begin{center}
        \begin{subfigure}[b]{0.49\textwidth}
            \centering
            \includegraphics[width=\textwidth]{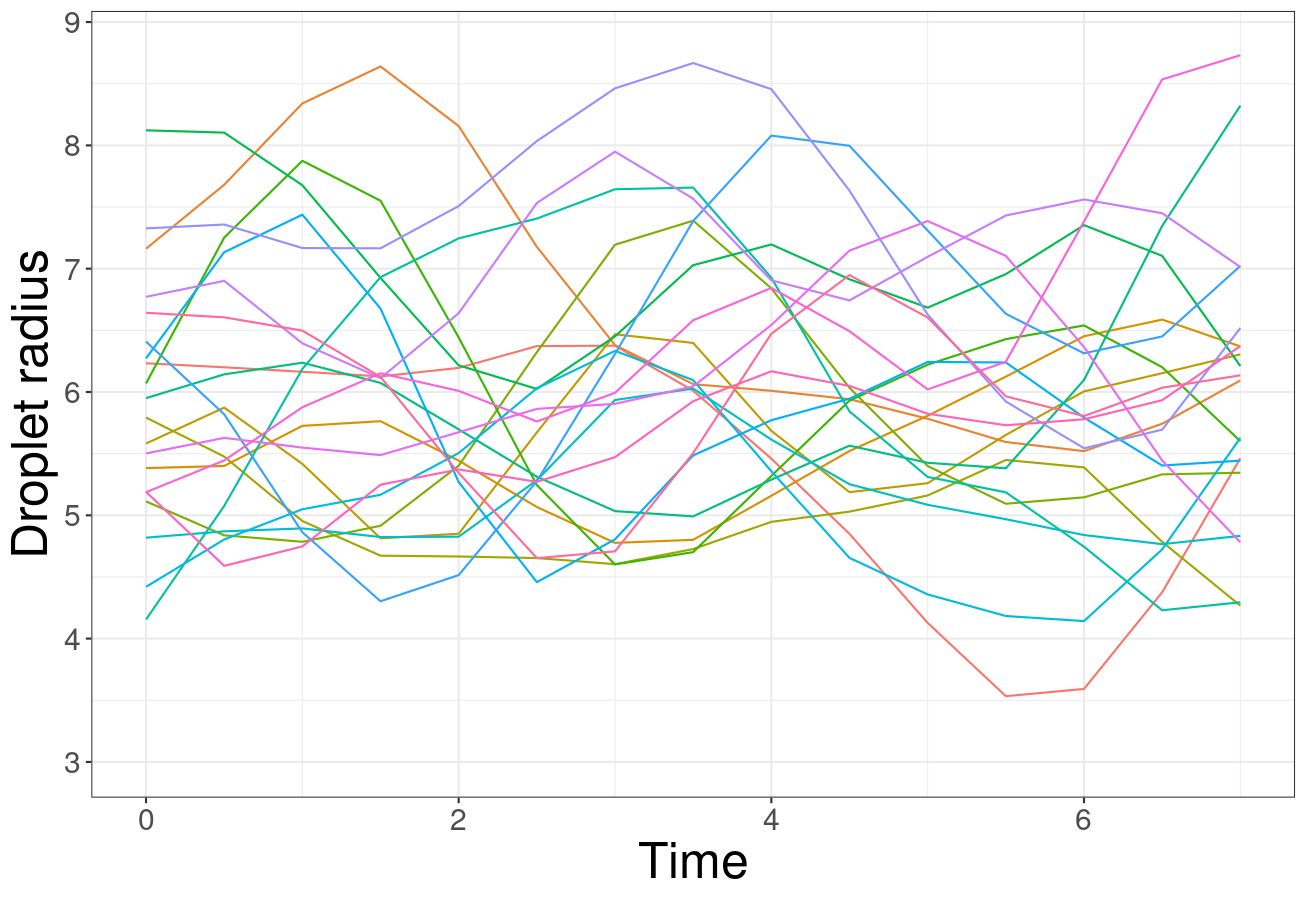}
            \caption{Prior draws}
            \label{fig:drop_prior}
        \end{subfigure}
        \hfill
        \begin{subfigure}[b]{0.49\textwidth}
            \centering
            \includegraphics[width=.92\textwidth]{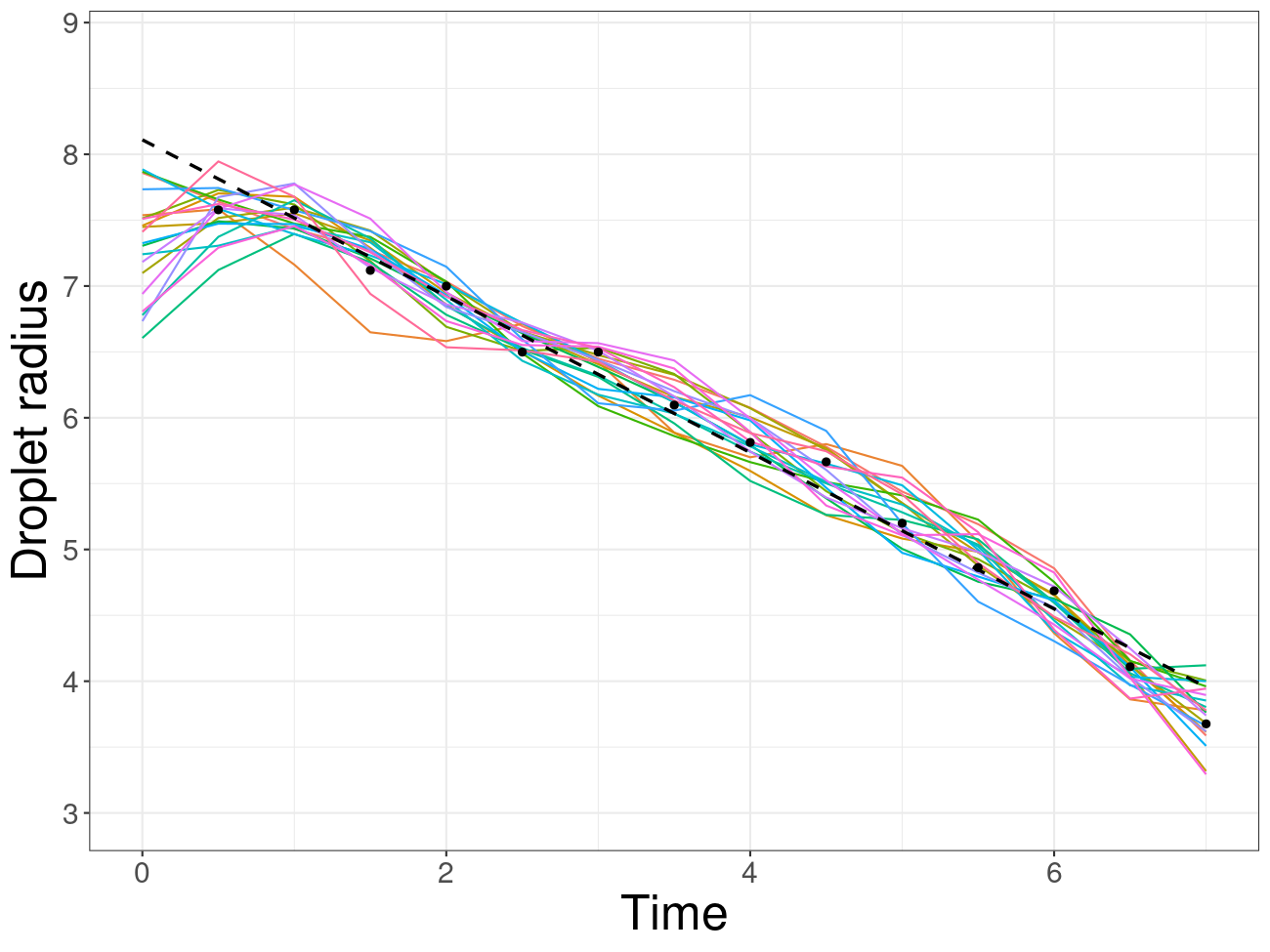}
            \caption{Posterior draws}
            \label{fig:drop_post}
        \end{subfigure}
        \caption{GP draws of the \textbf{(a)} prior and \textbf{(b)} posterior for the water droplet data. The colored curves represent each draw, the black dots are the observed data and the dashed line is the least squares estimate of the linear model.}
        \label{fig:drop_gp}
    \end{center}
\end{figure}

We use the GP in \autoref{sct:gp} to model the data, with the following prior settings:
$$
    \sigma^2 = 0.01, \quad m(t) = \frac{3+9}{2} = 6, \forall t \in \mathcal{X}, \quad K(t_1, t_2) = \exp\left\{ -\frac{1}{2}||t_1 - t_2||_2 \right\}, (t_1, t_2) \subset \mathcal{X}.
$$
As shown in \autoref{fig:drop_prior}, this choice leads to functions that obey the 3-9 micrometers restriction without becoming too restrictive as a consequence. In \autoref{fig:drop_post}, we observe that the posterior draws resemble a linear model except on $t = 0$, due to the missing observation.

In \autoref{tab:drop_fbst1}, we present the e-value for both hypotheses of interest assuming that $\mathcal{X}$ is either finite or infinite. Since small e-values provide strong evidence against $H_0$, with $\alpha = 0.05$ we conclude that both $H_0^1$ and $H_0^2$ should be rejected, i.e., Fick's law would fail.
\begin{table}[ht]
    \centering
    \caption{e-value of $H_0$ under finite and infinite $\mathcal{X}$ for the water droplet experiment.}
    \begin{tabular}{l|cc}
        \hline
        \multicolumn{1}{c}{}&\multicolumn{2}{c}{\textbf{Hypothesis}}\\
        \textbf{Assumption on} $\boldsymbol{t}$ & $H_0^1: g(t) = \beta_0 + \beta_1 t$ & $H_0^2: g(t) = \beta_0$\\
        \hline
        $t$ is discrete and finite & 0.0446 & 0\\
        $t$ is continuous & 0.0068 & 0\\
        \hline
    \end{tabular}
    \label{tab:drop_fbst1}
\end{table}

While this analysis shows that Fick's law is not exactly valid, it might still provide an adequate approximation, motivating the use of pragmatic hypotheses. This requires setting the threshold $\varepsilon$, which is detailed below.

In the original experiment, the radius of the droplets was obtained indirectly through Stoke's law, that is,
\begin{equation}
    V_T(t) = \frac{g(t)^2}{K_s} \Longrightarrow g(t) = \sqrt{V_T(t) \times K_s},
    \label{eq:stokes}
\end{equation}
where $V_T$ is the terminal velocity and $K_s = 8.446$. Since the mean velocity ($V_M$) was used in \eqref{eq:stokes} instead of $V_T$, there are two sources of measurement error: the estimate of $V_M$ (maximum error of $\delta = 0.14$) and switching $V_T$ for $V_M$ in \eqref{eq:stokes} (maximum error of $\eta = 0.3555$ \citep[][Example 1.3]{lassance2025}). We conclude that the margin of error of the radius is
\begin{align*}
    \epsilon :&= \max_{t \in T}|g(t) - y(t)| = \max_{t \in T}\left\{\left|\sqrt{K_sV_T(t)} - y(t)\right|\right\}\\
    &= \max_{t \in T}\left\{\left|\sqrt{K_s(V_M(t) - \delta - \eta)} - y(t)\right|, \left|\sqrt{K_s(V_M(t) + \delta + \eta)} - y(t)\right|\right\} \approx 0.6218.
\end{align*}
While $\epsilon$ relates to the $l_\infty$ distance, Lemma \ref{lem:lm_inf} uses the $l_2$ distance. To obtain an estimate of the latter from the former, we use Proposition 6.11 of \citet{folland2013}, which implies that
$$
    \sqrt{\frac{1}{n}\sum_{i = 1}^n(y(\boldsymbol{x}_i) - g(\boldsymbol{x}_i))^2} \le \sqrt{\frac{1}{n}}\epsilon \Longrightarrow \max_{i \in \{1, 2, \cdots, n\}}|y(\boldsymbol{x}_i) - g(\boldsymbol{x}_i)| \le \epsilon,
$$
thus $\varepsilon \approx 0.6218/\sqrt{15} \approx 0.1606$.

\autoref{tab:drop_fbst2} presents the e-values for the pragmatic hypotheses $Pg(H_0^i, d, 0.1606)$, $i \in \{1,2\}$. We assume either that $\mathcal{X} = \{0, 0.5, 1, \cdots, 7\}$ (original setting, discrete uniform) or that $t|P \sim P$, with $P \sim DP(1, U(0,7))$ (continuous uniform as centering distribution). Contrary to \autoref{tab:drop_fbst1}, the first hypothesis is not rejected, demonstrating that Fick's law provides a good approximation of the phenomenon.
\begin{table}[ht]
    \centering
    \caption{e-value of $Pg(H_0, d, 0.1606)$ under finite and infinite $\mathcal{X}$ for the water droplet experiment.}
    \begin{tabular}{l|cc}
        \hline
        \multicolumn{1}{c}{}&\multicolumn{2}{c}{\textbf{Original hypothesis}}\\
        \textbf{Assumption on} $\boldsymbol{t}$ & $H_0^1: g(t) = \beta_0 + \beta_1 t$ & $H_0^2: g(t) = \beta_0$\\
        \hline
        $t \in \{0, 0.5, 1, \cdots, 7\}$ & 1 & 0\\
        $t|P \sim P, \quad P \sim DP(1, U(0, 7))$ & 1 & 0\\
        \hline
    \end{tabular}
    \label{tab:drop_fbst2}
\end{table}

\section{Discussion}
\label{sct:disc}

Regarding the results of the application (\autoref{sct:droplet}), we believe to have demonstrated the importance of using pragmatic hypotheses whenever reasonable. While choosing $\varepsilon$ is not a simple task in nonparametric settings, there are strategies available for deriving it \citep{lassance2025}. Furthermore, while the e-value is not a measure of evidence against $H_0^c$ \citep{pereira2008}, combining it with a pragmatic hypothesis allows one to perform the Generalized FBST (GFBST, \citep{esteves2023}), which can discriminate ``evidence of absence'' from ``absence of evidence'' along with many other desirable properties. 

One of the main limitations of this work is in the strategy of performing variable selection. While the aforementioned GFBST allows for multiple testing without the necessity of correcting $\alpha$, variable selection is only possible through $\autoref{eq:lin_hyp}$ if the linear model hypothesis is not rejected. Therefore, one future research direction is developing tests that evaluate conditional independence without assuming a specific functional form for the relationship between variables.




\vspace{6pt} 




\authorcontributions{Conceptualization, all authors; methodology, R.F.L. Lassance and R.B. Stern; software, R.F.L. Lassance; validation, R.F.L. Lassance; formal analysis, R.F.L. Lassance; investigation, J.M. Stern; resources, R.F.L. Lassance; data curation, R.F.L. Lassance; writing---original draft preparation, R.F.L. Lassance; writing---review and editing, all authors; visualization, R.F.L. Lassance; supervision, R.B. Stern and J.M.Stern; project administration, R.B. Stern; funding acquisition, Interinstitutional Graduate Program in Statistics UFSCar-USP. All authors have read and agreed to the published version of the manuscript.}

\funding{This study was financed in part by the Coordenação de Aperfeiçoamento de Pessoal de Nível Superior - Brasil (CAPES) - Finance Code 001. This research was funded by FAPESP (grants 2019/11321-9 and CEPID CeMEAI 2013/07375-0) and CNPq (grants 309607/2020-5, 422705/2021-7 and PQ 303290/2021-8).}

\dataavailability{The data and the functions used in this study are available in GitHub at \url{https://github.com/rflassance/lmFBST}. These data were derived from the following resource available in the public domain: \url{https://scholarsmine.mst.edu/cgi/viewcontent.cgi?params=/context/masters_theses/article/6294} (page 42, accessed: 05/15/2024).}

\conflictsofinterest{The authors declare no conflicts of interest. The funders had no role in the design of the study; in the collection, analyses, or interpretation of data; in the writing of the manuscript; or in the decision to publish the results.}




\abbreviations{Abbreviations}{
The following abbreviations are used in this manuscript:\\

\noindent 
\begin{tabular}{@{}ll}
FBST & Full Bayesian Significance Test\\
GFBST & Generalized Full Bayesian Significance Test\\
GP & Gaussian Process\\
HPD & Highest Posterior Density\\
WRSS & Weighted Residual Sum of Squares\\
\end{tabular}
}

\appendixtitles{no} 
\appendixstart
\appendix
\section[\appendixname~\thesection]{Proofs}
\label{sct:proof}
\begin{proof}[Proof of \autoref{thm:lm_hpd}]
    The proof is done in parts:
    \begin{subproof}[Finite $\mathcal{X}$]
        Since $\mathcal{X}$ is finite, \autoref{eq:gp_ci} is the HPD. Therefore, if $\exists \boldsymbol{\beta} \in \mathbb{R}^k$ such that
        \begin{equation}
            \label{eq:fin_cond}
            (\boldsymbol{b}(\mathcal{X})\boldsymbol{\beta} - \mu(\mathcal{X}))' \Sigma(\mathcal{X}, \mathcal{X})^{-1} (\boldsymbol{b}(\mathcal{X})\boldsymbol{\beta} - \mu(\mathcal{X})) \le q_{(1-\alpha)}(\chi^2_{|\mathcal{X}|}),
        \end{equation}
        then the FBST does not reject $H_0$. Derivating the left side of \eqref{eq:fin_cond} in terms of $\boldsymbol{\beta}$, we observe that $\hat{\boldsymbol{\beta}}$ minimizes such expression.
    \end{subproof}
    
    \begin{subproof}[Infinite $\mathcal{X}$]
        In this case, the FBST does not reject the hypothesis iff $\exists \boldsymbol{\beta} \in \mathbb{R}^k$ such that $WRSS(\boldsymbol{b} \times \boldsymbol{\beta}) \le c_\alpha$. This is equivalent to not rejecting $H_0$ iff
        $$
            \left(\boldsymbol{y} - \boldsymbol{b}(\boldsymbol{X})\left(\boldsymbol{b}(\boldsymbol{X})'\boldsymbol{b}(\boldsymbol{X})\right)^{-1}\boldsymbol{b}(\boldsymbol{X})'\boldsymbol{y}\right)'\left(\boldsymbol{y} - \boldsymbol{b}(\boldsymbol{X})\left(\boldsymbol{b}(\boldsymbol{X})'\boldsymbol{b}(\boldsymbol{X})\right)^{-1}\boldsymbol{b}(\boldsymbol{X})'\boldsymbol{y}\right) = \boldsymbol{y}'\boldsymbol{M}\boldsymbol{y} \le c_\alpha,
        $$
        since $\left(\boldsymbol{b}(\boldsymbol{X})'\boldsymbol{b}(\boldsymbol{X})\right)^{-1}\boldsymbol{b}(\boldsymbol{X})'\boldsymbol{y}$ is the least squares estimate of $\boldsymbol{\beta}$.
    \end{subproof}
\end{proof}

\begin{proof}[\textbf{Proof of Lemma \ref{lem:lm_inf}}]
    The proof is found in \citet[][Theorem 2]{lassance2025}. \end{proof}

\begin{proof}[Proof of \autoref{thm:pg_hpd}]
    The proof is done in parts:
    \begin{subproof}[Finite $\mathcal{X}$] Lemma \ref{lem:lm_inf} implies that
    $$
        \tilde{\boldsymbol{\beta}}_h = \left(\boldsymbol{b}(\mathcal{X})'D_{\mathbb{P}(\mathcal{X})}\boldsymbol{b}(\mathcal{X})\right)\boldsymbol{b}(\mathcal{X})'D_{\mathbb{P}(\mathcal{X})}h(\mathcal{X}),
    $$
    and thus
    \begin{align*}
        d(H_0, h(\mathcal{X})) &= \sqrt{\sum_{\boldsymbol{x} \in \mathcal{X}} \mathbb{P}(\boldsymbol{x})\left(h(\boldsymbol{x}) - \boldsymbol{b}(\boldsymbol{x})'\tilde{\boldsymbol{\beta}}_h\right)^2}\\
        &= \sqrt{(h(\mathcal{X}) - \boldsymbol{b}(\mathcal{X})\tilde{\boldsymbol{\beta}}_h)'D_{\mathbb{P}(\mathcal{X})}(h(\mathcal{X}) - \boldsymbol{b}(\mathcal{X})\tilde{\boldsymbol{\beta}}_h)} = \sqrt{h(\mathcal{X})'\boldsymbol{N}h(\mathcal{X})} \le \varepsilon.
    \end{align*}
    Since the HPD is given by \autoref{eq:gp_ci}, the FBST does not reject $Pg(H_0)$ if and only if
    \begin{equation}
        \left\{
        \begin{array}{l}
             h \in \mathbb{H}: d(H_0, h(\mathcal{X})) = \sqrt{h(\mathcal{X})'\boldsymbol{N}h(\mathcal{X})} \le \varepsilon\\
             h \in \mathbb{H}: (h(\boldsymbol{X}) - \mu(\mathcal{X}))' \Sigma(\mathcal{X},\mathcal{X})^{-1} (h(\mathcal{X}) - \mu(\mathcal{X})) \le q_{1 - \alpha}(\chi^2_{|\mathcal{X}|}) 
        \end{array}
        \right.
    \end{equation}
    are intersecting ellipsoids. From Proposition 2 of \citet{gilitschenski2012}, the ellipsoids intersect if and only if
    $$
        \exists s \in (0,1) : \quad 1 - \mu(\mathcal{X})'\left(\frac{\varepsilon^2}{1-s}\boldsymbol{N}^{-1} + \frac{1}{s}\Sigma(\mathcal{X}, \mathcal{X})q_{(1 - \alpha)}(\chi^2_{|\mathcal{X}|}) \right)\mu(\mathcal{X})< 0.
    $$
    \end{subproof}

    \begin{subproof}[Infinite $\mathcal{X}$]
    The FBST does not reject $Pg(H_0)$ if
    {\small
    \begin{equation}
        \label{eq:fbst_equiv}
        \inf_{h \in HPD}d(H_0, h)^2 = \inf_{h \in HPD} \inf_{h_0 \in H_0} d(h_0, h)^2 = \inf_{h_0 \in H_0} \inf_{h \in HPD} \int_{\mathcal{X}} (h_0(\boldsymbol{x}) - h(\boldsymbol{x}))^2 d\mathbb{P}_{\boldsymbol{X}}(\boldsymbol{x})\le \varepsilon^2.
    \end{equation}
    }
    The ellipsoid $G := \{\boldsymbol{z} \in \mathbb{R}^{n^*}: (\boldsymbol{z} - \overline{\boldsymbol{y}}_{\boldsymbol{X}^*})'D_{n_{\boldsymbol{X}^*}}(\boldsymbol{z} - \overline{\boldsymbol{y}}_{\boldsymbol{X}^*}) \le c_\alpha\}$ is such that, for any function $h(\cdot)$ where $\exists \boldsymbol{z} \in G: h(\boldsymbol{X}^*) = \boldsymbol{z}$, we can conclude that $h \in HPD$. Therefore, the $HPD$ contains functions that are linear outside of $\boldsymbol{X}^*$, and thus
    \begin{align*}
        \inf_{h_0 \in H_0}\inf_{h \in HPD} \int_{\mathcal{X}} (h_0 - h(\boldsymbol{x}))^2 d\mathbb{P}_{\boldsymbol{X}}(\boldsymbol{x}) =& \inf_{h_0 \in H_0}\inf_{\boldsymbol{z} \in G}\sum_{i = 1}^{n^*} (\boldsymbol{z}_i - h_0(\boldsymbol{X}_{i,.}^*))^2\mathbb{P}(\boldsymbol{X}_{i,.}^*)\\
        =& \inf_{\boldsymbol{z} \in G}\inf_{h_0 \in H_0} (\boldsymbol{z} - h_0(\boldsymbol{X}^*))'D_{\mathbb{P}(X^*)}(\boldsymbol{z} - h_0(\boldsymbol{X}^*))\\
        =& \inf_{\boldsymbol{z} \in G}(\boldsymbol{z} - b(\boldsymbol{X}^*)\hat{\boldsymbol{\beta}}_{\boldsymbol{z}})'D_{\mathbb{P}(X^*)}(\boldsymbol{z} - b(\boldsymbol{X}^*)\hat{\boldsymbol{\beta}}_{\boldsymbol{z}}),
    \end{align*}
    where $\hat{\boldsymbol{\beta}}_{\boldsymbol{z}} = (b(\boldsymbol{X}^*)'D_{\mathbb{P}(X^*)}b(\boldsymbol{X}^*))^{-1}b(\boldsymbol{X}^*)'D_{\mathbb{P}(X^*)} \boldsymbol{z}$, thus $\inf_{h \in HPD}d(H_0, h)^2 = \inf_{\boldsymbol{z} \in G}\boldsymbol{z}'\boldsymbol{M}\boldsymbol{z}$. Therefore, the FBST does not reject $H_0$ if the ellipsoids $G$ and $\{\boldsymbol{z} \in \mathbb{R}^{n^*}:\boldsymbol{z}'\boldsymbol{M}\boldsymbol{z} \le \varepsilon\}$ intersect, which can be verified through Proposition 2 of \citet{gilitschenski2012}.
    \end{subproof}
\end{proof}



\begin{adjustwidth}{-\extralength}{0cm}

\reftitle{References}


\bibliography{ref.bib}

\PublishersNote{}
\end{adjustwidth}
\end{document}